\documentclass[prodmode,acmtomm]{acmsmall} 

\usepackage[ruled]{algorithm2e}
\usepackage{caption}
\usepackage{calc}
\usepackage{tabularx,ragged2e}
\usepackage{array,booktabs}

\usepackage[ruled]{algorithm2e}
\renewcommand{\algorithmcfname}{ALGORITHM}
\SetAlFnt{\small}
\SetAlCapFnt{\small}
\SetAlCapNameFnt{\small}
\SetAlCapHSkip{0pt}
\IncMargin{-\parindent}

\acmVolume{9}
\acmNumber{4}
\acmArticle{39}
\acmYear{2016}
\acmMonth{3}

\doi{0000001.0000001}

\usepackage{array}
\newcolumntype{M}{>{\centering\arraybackslash}m{\dimexpr.25\linewidth-12\tabcolsep}}

\renewcommand{\algorithmcfname}{ALGORITHM}
\SetAlFnt{\small}
\SetAlCapFnt{\small}
\SetAlCapNameFnt{\small}
\SetAlCapHSkip{0pt}
\IncMargin{-\parindent}

\acmVolume{0}
\acmNumber{0}
\acmArticle{0}
\acmYear{2016}
\acmMonth{0}

\begin{document}


\markboth{Nilsson et al.}{
Applying Seamful Design in Location-based Mobile Museum Applications
}

\title{
\textcolor{black}{Applying Seamful Design in Location-based Mobile Museum Applications}
}
\author{Tommy Nilsson\textsuperscript{1}
\affil{University of Nottingham}
Carl Hogsden\textsuperscript{2}
\affil{University of Cambridge}
Charith Perera\textsuperscript{3}
\affil{The Open University}
Saeed Aghaee\textsuperscript{4}
\affil{University of Cambridge}
David J. Scruton\textsuperscript{5}
\affil{University of Cambridge}
Andreas Lund\textsuperscript{6}
\affil{Ume\r{a} University}
Alan F. Blackwell\textsuperscript{7}
\affil{University of Cambridge} \vspace{-15pt}}

\begin{abstract}

\textcolor{black}{The application of mobile computing is currently altering patterns of our behavior to a greater degree than perhaps any other invention. In combination with the introduction of power efficient wireless communication technologies, such as Bluetooth Low Energy (BLE), designers are today increasingly empowered to shape the way we interact with our physical surroundings and thus build entirely new experiences. 
However, our evaluations of BLE and its abilities to facilitate mobile location-based experiences in public environments revealed a number of potential problems. Most notably, the position and orientation of the user in combination with various environmental factors, such as crowds of people traversing the space, were found to cause major fluctuations of the received BLE signal strength. These issues are rendering a seamless functioning of any location-based application practically impossible. Instead of achieving seamlessness by eliminating these technical issues, we thus choose to advocate the use of a seamful approach, i.e. to reveal and exploit these problems and turn them into a part of the actual experience. In order to demonstrate the viability of this approach, we designed, implemented and evaluated the \textit{Ghost Detector} - an educational location-based museum game for children. By presenting a qualitative evaluation of this game and by motivating our design decisions, this paper provides insight into some of the challenges and possible solutions connected to the process of developing location-based BLE-enabled experiences for public cultural spaces.} 

\end{abstract}

\category{J.0}{Computer Applications}{General}
\category{H.5.2}{Information Systems-Information Interfaces and Presentation}{User Interface}

\terms{Design, Human Factors, Experimentation}

\acmformat{Tommy Nilsson, Carl Hogsden, Charith Perera, Saeed Aghaee, David Scruton, Andreas Lund,
and Alan F. Blackwell, 2016. Applying Seamful Design in Location-based Mobile Museum Applications.}




\maketitle

\begin{bottomstuff}
The project was supported using public funding by Arts Council England. Tommy Nilsson's work was supported by the Engineering and Physical Sciences Research Council (project EP/N014243/1). Dr. Saeed Aghaee's work was supported by a Swiss National Science Foundation Early Postdoc Mobility fellowship (P2TIP2 152264). Dr. Charith Perera's work was supported by CECS Deans Travel Grant, International Alliance of Research Universities (IARU) Travel Grant, and The ANU VC Travel Grant. 
Author's addresses: Tommy Nilsson, The Mixed Reality Laboratory, Nottingham, NG8 1BB, UK; Charith Perera, Department of Computing, The Open University, Walton Hall, Milton Keynes, MK7 6AA, UK; Carl Hogsden and David J. Scruton, The Fitzwilliam Museum, Trumpington Street, Cambridge CB2 1RB, UK; Andreas Lund, Department of Informatics, Umea University, SE-901 87 Umea, Sweden; Saeed Aghaee and Alan F. Blackwell, William Gates Building, 15 JJ Thomson Ave, Cambridge CB3 0F, UK;
emails: 
psxtn2@nottingham.ac.uk\textsuperscript{1}, carl@hogsden.org\textsuperscript{2}, charith.perera@ieee.org\textsuperscript{3}, saeed.aghaee@cl.cam.ac.uk\textsuperscript{4}, djs94@cam.ac.uk\textsuperscript{5}, alund@informatik.umu.se\textsuperscript{6}, afb21@cam.ac.uk\textsuperscript{7}
\end{bottomstuff}

\section{Introduction}

The pioneer ubiquitous computing researcher Mark Weiser famously envisioned a future in which whole environments would be enhanced through computational resources providing information and services whenever and wherever desired, a world in which immersion with digital information would become a permanent state of our lives \cite{weiser1991}. After long being hampered by the nature of desktop computers, which restricts all digital interactions to a relatively constrained time and space, the current proliferation of mobile devices makes such ideas sound increasingly realistic. Close to 70\% of the British population now owns a smartphone \cite{styles2013} and research indicates that their popularity is indeed in the process of surpassing that of the traditional desktop computer \cite{busching2012}.

Although in many ways replacing traditional computers, it would be a mistake to think of modern mobile devices simply as smaller versions of their desktop counterparts. The development of mobile cameras and sensors, such as QR readers, gyroscopes or radio signal receivers, has granted the contemporary smartphone a substantial degree of context awareness. Whereas traditional desktop computers had to rely predominantly on data contained on their hard drives, our smartphones are increasingly reliant and responsive to information retrieved directly from our physical surroundings, which is in turn enabling a wide variety of interactive modalities \cite{she2014}. Consequently a growing number of people are now using their phones to enjoy various ubiquitous location-based services (LBS) \cite{liu2013}.

Correspondingly, a growing number of sectors is attempting to exploit this emerging technological landscape by using physical objects and places as communication channels to reach mobile users. Essentially any physical artifact equipped with an identifier and wireless connectivity can today be incorporated into this communication network, sometimes referred to as the Internet of Things, or simply IoT \cite{golding2011}. The idea of IoT has however its fair share of flaws. Until recently, perhaps the most prominent one came in the form of high power consumption of wirelessly accessible information services, forcing many stakeholders to instead adopt cumbersome and often somewhat unaesthetic interfaces, such as QR code scanners. 

Recent advances in sensors and ultra low power wireless data transfer technologies, such as the Bluetooth Low Energy standard (henceforth BLE), are however contributing towards a gradual erosion of this limitation. Unlike its Bluetooth predecessor, BLE has a significantly reduced power consumption, with multiple sources claiming that a BLE beacon can operate continuously for months on a single coin battery \cite{kamath2012}. In combination with a low cost, these factors could very well establish BLE as the dominant technology in the practice of granting physical objects a wireless connectivity. BLE beacons can thus be used to relatively quickly and efficiently augment any environment through a degree of ambient intelligence. By communicating with users through their mobile devices, such smart environments can then effectively provide them with information related to their current situation and position in space. Given that BLE is expected to be supported by billions of mobile devices in the near future\cite{west2013} and achieve a strong presence in several market segments \cite{gomez2012}, it seems inevitable that similar experiences are about to play an increasingly important role in our lives. 

And so, although innovations such as BLE are now slowly turning the dreams of Mark Weiser into reality, we are still only beginning to understand the possibilities offered by this technology and the implications a constant immersion and interconnection through computation might have. These innovations and their ubiquity are inevitably pressing us towards rethinking the way we go about designing everyday experiences. There is a clear need to generate a better understanding of technologies such as BLE in order to support future practitioners. 

Likewise, we ought to consider a range of relevant human factors as well. Before the presence of such ubiquitous systems expands into contexts and situations traditionally untouched by computing technologies, we would do well to pause and contemplate whether we are ready to embrace them. History shows us that a technological solution will not become successful merely because it is working, but rather because a sufficient number of users is willing to accept it. This willingness of the general public to accept and use a new interaction paradigm on a daily basis ought to be called into question. Why should a person, used to be going about daily life without the assistance of any ubiquitous information system now decide to start taking advantage of one? Early adopters and other technology enthusiasts visiting places such as Apple stores \cite{apple2013} are one thing, but the general public is quite another. Accessibility to non-digital natives, our sense of privacy when interacting with smart public spaces and the overall enjoyment are but a few of the factors that will inevitably have a decisive impact on the holistic user experience. The ability to comply with the relevant human needs will be crucial for the success of any truly ubiquitous information system. 

Against the backdrop of this emerging technological landscape, the goals of the study to be presented in this paper have been set as follows: 

\begin{itemize}
\item By evaluating the performance of BLE technology in public environments, we will seek to generate insight into the challenges as well as the opportunities associated with the use of this technology. 

\item This preliminary investigation of BLE technology shall be translated into a conceptual design of a solution that would best address these challenges. 

\item Finally, as a proof of concept, a prototype of a ubiquitous information service shall be developed and implemented into a group of museum exhibitions in order to help us assess the viability of our design approach and its implications on the user experience.
\end{itemize}

Reaching these goals is not only expected to produce a set of guidelines for future designers and developers dealing with mobile devices and BLE, but also provide insights contributing to the ongoing theoretical discourse on ubiquitous computing and its possible roles in improving our daily lives. Moreover, as will be argued, studies of BLE are still scarce and focus predominantly on technical aspects. By exploring the use of BLE-enabled systems in the context of public environments and by attempting to solve some of its inherent technical problems through a design framework rather than by improving the technology, it is our firm belief that this paper will fill a gap in the existing discourse and aid practitioners developing context-aware mobile applications, as well as architects and other planners seeking to establish built environments that would best act as entities that engage and interact with people in specific situations. 

A brief overview of the current state of relevant research will be given in section 2. Our project and its objectives will be presented in more detail in section 3. Our preliminary investigation of BLE technology in the context of public environments will be presented in section 4. In section 5 we propose a suitable design approach. A set of usability evaluations of a prototype system is presented in section 6. A discussion follows in section 7, before the paper concludes in section 8. 

\section{Related Work}
The design, development and evaluation of novel ubicomp solutions has attracted a substantial level of interest over the past years. Traditional cultural environments, such as museums, have not been exempted from this trend. \cite{jacobs2015} \cite{vancat2012} For instance Gonz{\'a}lez, Organero and Kloos \cite{gonzales2008} attempted to find the best way of utilizing ubiquitous computing to develop an infrastructure for innovative learning spaces. They concluded that, much due to their versatility, mobile phones will in the near future be able to support all the different types of processes needed in learning scenarios.


\textcolor{black}{Today a range of solutions, such as WiFi or Global Positioning System (GPS), allows mobile users to interact with physical spaces based on their location. There is however a particularly strong case to be made for BLE as a preferable solution for public cultural spaces, such as museum exhibitions. For instance, an analysis of the performance of BLE in indoor positioning applications, carried out by Faragher and Harle \cite{Faragher2014b}, found that even a relatively sparse deployment of BLE beacons results in a significant positioning improvement compared to existing WiFi infrastructures. Specifically, the study found that the positioning error of a WiFi tracking scheme was less than 8.5 m 95\% of the time, while a network of BLE beacons (broadcasting at a frequency of 10 Hz and power levels around -20 dBm) achieved an accuracy of less than 2.6 m 95\% of the time. This level of precision cannot currently be matched by any other mainstream positioning technology, including the GPS. \cite{Sterling2014} Moreover, a study by Wein \cite{wein2014}, evaluating the preferences of museum visitors, suggests that interfaces leaving a visible footprint on the environment, such as QR codes, can be perceived as having a negative impact on the content and appearance of the museum space, thus harming the user experience.}

\textcolor{black}{In line with the findings of Wein, we sought to create an alternative museum experience while leaving the original space intact, rather than contaminating it with QR codes, NFC tags and other tools. The limited size of BLE beacons made it feasible to store them within a drawer or underneath a shelf, thus allowing us to simply enhance selected locations in a museum space through an invisible layer of information accessible through mobile devices.}

\textcolor{black}{Previous research has however also identified a number of potential challenges associated with the use of BLE technology.} Another study by Faragher and Harle \cite{faragher2015} revealed that the comparatively low bandwidth of BLE makes it more likely to fade fast, resulting in considerably more fluctuations in received signal strength (henceforth RSS) than with WiFi. They conclude that achieving accurate positioning is highly problematic unless we learn to understand and deal with these large fluctuations. 

Albeit somewhat discouraging, these findings were not unexpected. Ubiquitous computing is based on the principle of multiple distinct devices providing highly dispersed input, output and computational capabilities. Practically every such system, no matter how carefully engineered, will always end up being a subject to variables beyond our control and as such cannot be expected to operate perfectly consistently. Due to the extreme diversity of environments where they are expected to operate, mobile technologies are arguably suffering by these technical limitations to an even greater degree than other ubicomp systems. Patchy network coverage, fluctuating signal strength and deviations in positioning are but a small sample of the issues mobile users had to grow used to. 

These deviations and discontinuities between what a mobile device observes and what actually happens, are by some researchers described as \textit{seams} \cite{weiser1993}. Broll and Benford \cite{benford2005} argue that seams are traditionally seen as a detracting element of user experience. As a result, developers are frequently putting significant effort into making systems as seamless as possible by eliminating or hiding any inconsistency. This is however often done at the cost of expensive investments into a better technology. 

In search of a more time and cost effective solution, Broll and Benford \cite{benford2005} refer to Mark Weiser, who endorsed \textit{invisibility} as the main design goal of ubiquitous computing.\cite{weiser1993} In Weiser’s view, designers should hide the complexity and infrastructure of tools to prevent them from intruding on user's consciousness. This will in turn allow users to focus on their tasks and not on the tool itself. As an example of such invisible technology, Weiser  names electricity, which we use, yet do not have to attend to. Interestingly, according to him, such invisibility should not automatically be seen as an equivalent to seamlessness. He warns that making things seamless amounts to reducing all system components to their lowest common denominator and making everything the same. In his view, such approach would sacrifice the richness of every tool just to achieve a bland compatibility. Instead, he suggests an alternative approach in the form of \textit{seamful systems with beautiful seams} as the goal designers should strive for. In other words, if properly designed, seamfully integrated parts of a system could still provide a seamless interaction. 

This view has been refined by Chalmers \cite{chalmers2003} who argues that integrating seams into an experience is hard, but the quality of interaction can be improved if we allow each component to be itself. In line with these arguments, Broll and Benford \cite{benford2005} propose a new paradigm in the form of a \textit{seamful design framework}, revolving around the idea of revealing and exploiting technical limitations in ubiquitous computing. Since many of the seams arising when using a technology are in fact inevitable, instead of trying to hide them, they argue that we ought to use them to enhance the actual experience. In fact most of us are already using seams to enhance our experience, often without even knowing about it. For instance, we have all been in situations where we received an unsolicited phone call and dismissed the caller by claiming that we're driving through a tunnel and the signal is getting too bad to talk. In such simple situations, we are in fact using the common knowledge of an inconsistent signal coverage to our own advantage. Same mindset, they argue, can be employed by designers when dealing with some of the intrinsic problems of ubiquitous computing. \textcolor{black}{In other words, a seamful design framework is centered around the idea of deliberately revealing seams to users, and taking advantage of features usually considered as negative or problematic.\cite{chalmers2003}} 

\textcolor{black}{Perhaps the best known example of a seamful mobile application was the 2001 chase game \textit{Can You See Me Now} (CYSMN).\cite{benford2006} In CYSMN a small number of \textit{runners} was tasked to move around through the streets of Sheffield. Meanwhile, a different group of players was introduced to a virtual representation of the same city online. Their position was tracked in real time and displayed to the runners through their PDA devices. Likewise, the position of runners was tracked using GPS and displayed to the online players in their virtual environment. The goal of the runners was then to catch as many of the online players as possible by moving into sufficient proximity of the position they were occupying in the virtual world. The fluctuating GPS coverage made it difficult for runners to catch online players in certain areas of the map and keeping track of GPS coverage became an important strategic advantage, thus adding an extra dimension to the gameplay.}

In another attempt to demonstrate the viability of a seamful design approach, Chalmers \cite{chalmers2003} developed a mobile location-based game called \textit{Bill}, that was centered around the idea of exploring seams in wireless networking. Players were encouraged to collect virtual coins distributed on predetermined GPS coordinates. When finding areas with stronger signal coverage, or so called \textit{access points}, players could upload their coins to the game server in exchange for credits. On the other hand by staying out of these network covered areas, players avoided pickpockets threatening to steal their coins. Due to signal strength fluctuations, positions of the access points were dynamic and keeping track of them became an important part of the game. Chalmers concluded that by exploiting seams (e.g. inconsistent network coverage and signal strength fluctuations), multiple aspects, including interaction, gameplay and usability were improved. 

Both Bill and CYSMN indicated that a seamful design approach can indeed produce positive results. None of the researchers mentioned above did however claim that seamlessness is always bad, or that seamfulness is always good. Rather, by highlighting this underexposed and underused design approach, they were merely attempting to demonstrate that there is a lot of room for seamful work. When looking at existing related research, it seems undeniable that an overwhelming majority of practitioners is indeed attempting to eliminate seams by assuming accuracy in location awareness \cite{yin2015}. One such example is a museum augmented reality guide developed as a joint project between the Louvre in France and the Japanese printing company Dai Nippon Printing \cite{Arnaudov2008} in an attempt to gain experience in innovative multimedia approaches. By equipping selected artifacts with RFID tags and distributing ultra mobile PCs to its visitors, the Louvre could provide every user with relevant commentary and animations triggered on the PCs simply by moving into the transmitting range of the RFID tag of respective artifacts. The perceived usability and overall experience was evaluated through questionnaires and observations of the users. The study found that people paid greater attention to the artifacts and acted more in accordance with the route guide than if relying on more traditional solutions, such as static PDA screens and audio commentaries. 

Due to low complexity of the technology, seamless design proved to work well in this particular case. The question we had to deal with in our project was however whether a ubiquitous BLE-enabled information system is likely to reach a level of complexity where seamless operation will become problematic and thus prompting us to explore a more seamful design approach instead.

Regardless of the design approach, the practice of implementing new media technology into traditional knowledge spaces has been generating predominantly positive results. Jones and Jo \cite{jones2004} studied the use of ubiquitous technology in learning environments and argued that it enables students to access education more naturally. They went as far as concluding that ubiquitous computing may in fact be the new hope for the future of education. 

\textcolor{black}{Similarly, studies have shown that the introduction of computing technology into museum environments can have a positive impact on the user experience, particularly so when it comes to engaging visitors in explanations of complex artifacts. \cite{yamazaki2009}}

Despite these findings, we have yet to see any widespread use of such technology in the context of everyday public spaces. It would be unwise to attribute this situation wholly to constraints such as cost of these systems. Research shows that the willingness to embrace a technological solution is always influenced by a multitude of variables. For instance, in 1989 Fred Davis pioneered the Technology Acceptance Model (TAM).\cite{davis1989} He argued that adoption of new technology depends predominantly on two external variables: perceived usefulness, i.e. the degree to which a person believes that using a particular system would enhance her performance, and perceived ease of use, i.e. the degree to which a person believes that using a particular system will be free from effort. Due to our solution being implemented in a museum environment, we had to anticipate a wide spectrum of users, including non-digital natives. Issues related to social acceptance thus had to be factored in throughout the development process as well. 

TAM was later expanded by Timothy Teo to accommodate for learning spaces, such as museums. \cite{teo2008}Teo claimed that perceived low complexity goes hand in hand with perceived ease of use. Demetriadis goes as far as pointing out that initial difficulties in using a technology can cancel out the potential benefits that users believe they will gain from its use \cite{Demetriadis2003}.

The models mentioned above provide us with an ideal framework through which we can estimate the potential for mainstream adoption of a BLE-enabled system. The aforementioned seamful design paradigm is on the other hand providing us with potentially important clues regarding the design approaches that are sufficiently time and cost effective to be applicable when planning and developing BLE-enabled systems for public environments. Indeed, as will be argued in the following sections, seamful design played an instrumental role in the design and development of our solution. 

\section{Problem Definition}
The University of Cambridge Museums (UCM) has a long-standing interest in researching the potential use of new media technology in enriching the way we engage with cultural information \cite{rosati2013}. These efforts provided us with an ideal setting to examine BLE and evaluate its potential in the context of public spaces. 


Specifically, the locations selected for our project were the Sedgwick Museum of Earth Sciences, Whipple Museum of the History of Science, and the Museum of Archaeology and Anthropology. These mid-size university owned museums are all located in central Cambridge within walking distance of each other. Their family friendly environments meant that all major age demographics, including children, can be found among their visitors. Since one of our goals was to find out how well BLE systems would perform in public everyday life environments, the nature of these museums made them into an excellent choice of testing ground.

One particular factor that makes the museum context stand out from most other public environments is the fact that most exhibitions contain educational elements of some sort. In order to comply with the museum setting, our goal was not to create a BLE-enabled experience that would merely happen to take place in a museum. Rather, the goal was to enhance the museum environment in a usable way by improving the existing experience. Since the museums serving as hosts to our study offered an experience that was educational in nature, learning and the possible role of computer technology in supporting education had to be factored in throughout the design process.

It is however vital to understand that although the design solution about to be described in this paper had to be adapted to the setting of particular museum exhibitions, our aim was not to develop an ad hoc application. Rather, we were concerned with highlight the problems and possible design solutions relevant to the process of integrating a BLE-enabled mobile system into public environments in general. As such the research conclusions brought forth aspire to be applicable to a broad spectrum of applications beyond just the museum setting. 


\section{Preliminary Investigation}
BLE was designed as a low energy IoT platform to enable an invisible and unobtrusive way of interacting with physical objects. One of our main concerns was however that the low power consumption would translate into shorter transmission range when compared to the traditional bluetooth technology. To methodologically assess the possible use of this technology, we therefore had to start out by examining the distance across which BLE beacons were able to establish a reasonably stable communication with mobile devices. 


In order to carry out such an experiment, we first had to develop a custom mobile application capable of measuring BLE signal strength. This ad hoc application was programmed to continuously scan the surroundings in search of any BLE beacon. Every time such a beacon was detected and a connection with it was successfully established, the application produced a numerical value indicating the RSS of the given beacon. The application was built using the Android Eclipse development environment and written in Java. 

\textcolor{black}{The prime objective of this evaluation was to provide us with a better understanding of how BLE beacons operate in their natural habitats out in the wild. Hence, it was imperative for the evaluation to be carried out in the setting of a real museum exhibition featuring many of the elements that are present in typical public spaces hosting cultural content, such as rooms surrounded with concrete walls and filled with wooden shelves or crowds of people constantly traversing the space. }

During the actual experiment, a single BLE beacon was placed in turn at different locations in the Sedgwick Museum of Earth Sciences along its main corridor at a height of approximately 1 meter above the ground. The beacon was set to emit signals at a rate of 10Hz and TXPower set to -4 dBm. To account for any abnormal fluctuations in the signal received from the beacon, besides RSS, our mobile application had to additionally be programmed to measure standard deviation (henceforth SD). Each SD value was calculated based on 50 RSS samples (i.e. the total number of signal pulses received over a duration of 5 seconds). The smartphone Samsung Galaxy S3, running Android 4.3, was used during the experiment. Data collection was carried out by walking through the entire museum space while measuring changes in the RSS and SD values (i.e. the dependent variables) in relation to our location in the museum space (i.e. the independent variable). Furthermore, in order to account for the possible effect that our facing direction could have on the received signal strength, each sample location in the museum space had the signal values measured while facing in turn north, west, south and east. Approximately every square meter of the museum space was measured using this procedure. Due to the relatively tedious and time consuming nature of this data collection process, five consecutive days had to be spent before RSS distribution of the entire museum space was measured. All of these sessions were carried out during the opening hours of the museum in order to accurately assess the performance of BLE in the dynamic conditions typically encountered in public environments (e.g. the high layout complexity in combination with people constantly traversing the space). The whole experimental study was conducted with full knowledge and consent of the Sedgwick museum staff. Since the study didn't directly involve any human participants (beyond the research team), no potentially ethically problematic issues were encountered.

\textcolor{black}{A map of sampling points (see figure 6) and a set of tables featuring all the RSS values collected across one of the museum sections (see table II) have been included in the appendix section of this document to provide an overview of the collected data. A few of our key findings are however in the need of a closer elaboration:}
\textcolor{black}{
\begin{itemize}
  \item As expected, the study showed a steady RSS deterioration whenever moving away from a BLE beacon. Our results do however suggest that under typical conditions a transmission power set to -4 dBm (400 microWatts) is generally sufficient to secure propagation of the signal across a minimum distance of roughly 5 meters in spite of the wooden shelves and other objects unevenly distributed throughout the museum space. This means that a network of relatively few BLE beacons is sufficient to cover most indoors environments, provided that they are adequately positioned. 
\item Beyond this safe range, our experiment uncovered substantial inconsistencies in the RSS falloff across space, as well as irregular fluctuations over time. The nature of the museum interior with large shelves and artifacts allocated throughout the space proved to be an important factor contributing to the somewhat unpredictable signal coverage. The museum visitors, seemingly randomly walking around in the space, were found to be another factor causing signal interference. In fact, on a  few occasions, larger crowds were able to block out the signal entirely. 
  \item The spatial orientation of the mobile user was also found to have a dramatic impact on the RSS. For instance, as seen in Figure 7, by turning our back towards the beacon, and thus effectively obstructing the path of the signal, the RSS could drop in some cases by as much as 10 dBm, which in this context is quite considerable. It thus seems reasonable to say that users' spatial orientation should be taken into account by any BLE-enabled location-based solution. 
\end{itemize}
} 

\begin{figure}[h!]
\begin{center}
\includegraphics[scale=0.18]{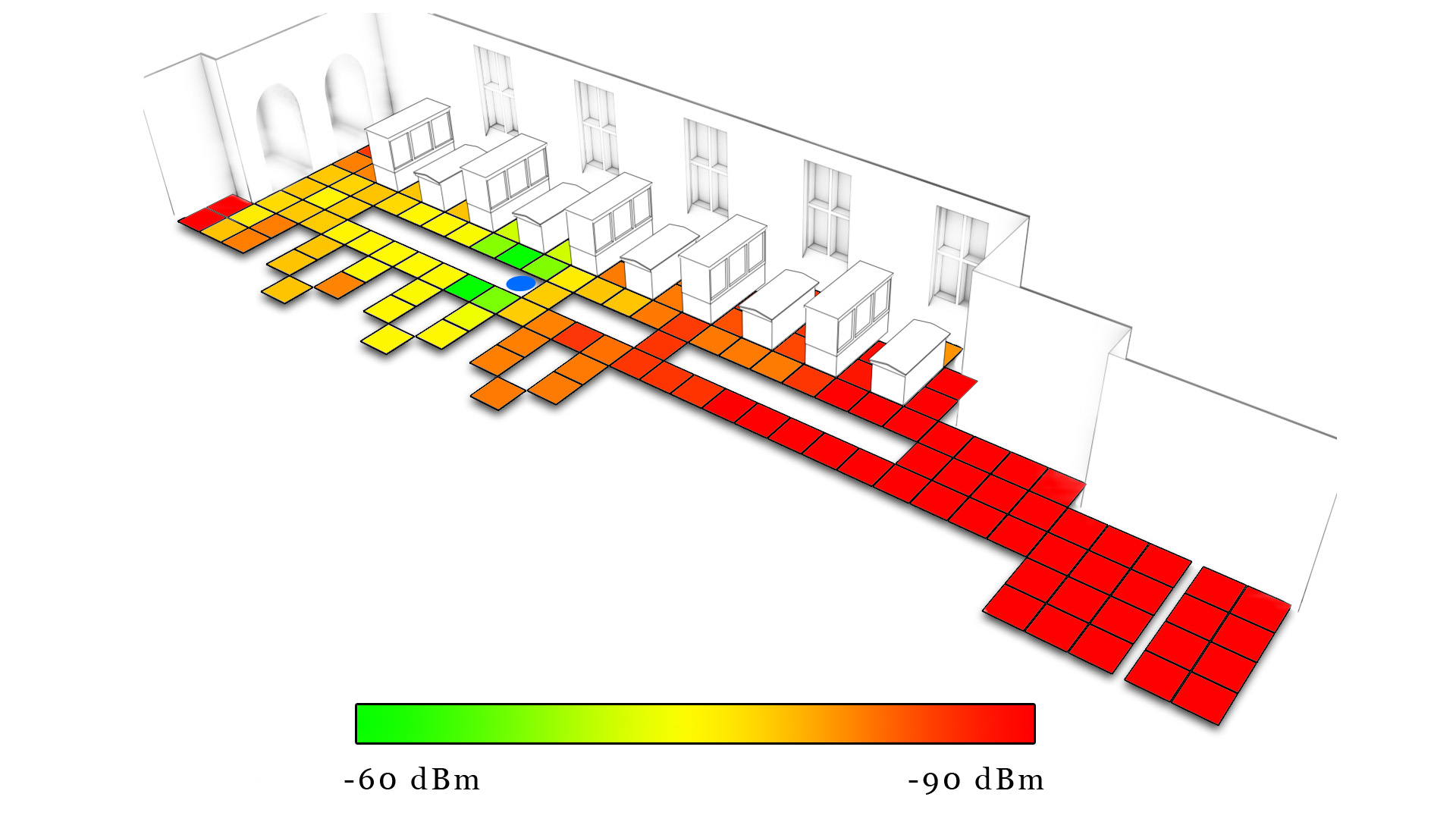}
\caption{Received signal strength was heavily dependent on a range of environmental factors. Location of the BLE beacon is represented by the blue dot. Every square on the map represents 1$m\textsuperscript{2}$.}
\end{center}
\vspace{-10pt}
\end{figure}

\textcolor{black}{All these signal strength inconsistencies made it clear to us that relying on BLE for accurate positioning would be unwise. Any attempt of determining absolute position of the user based on a set of RSS values would likely produce inaccurate results.} However, as has been discussed, the issues causing such uncertainty are an inseparable component of most public environments and as such must be factored in during the design process of any BLE-centered system intended for use in this context.\textcolor{black}{ The issue of unpredictable environmental factors and its effect on location based services will be addressed more in depth in the following section.} As will be explained, unconventional approaches, such as seamful design, might hold the key to overcoming many of these obstacles. 

\section{Smart Spaces: Conceptual Design}
As has been argued in the introductory section, our solution had to consist of two main components: firstly, a specific constellation of BLE beacons allocated in the museum space and secondly, a mobile application communicating with this network of beacons. 

Our technological evaluation has however manifested a number of trade-offs associated to the use of BLE technology. Most importantly, the precision and range across which mobile devices and BLE transmitters can communicate was found to be highly dependent on multiple variables beyond our control, making the design of any predictable experience problematic. These results left us with the seemingly intimidating task of coming up with a concept that would be operational in spite of these difficulties.

Although there might be a way to address many of the issues we came across by improving or reengineering the BLE technology, this would likely be a time and resource consuming process. In our situation it was seen as reasonable to instead first explore a seamful design approach which would allow us to take advantage of technical limitations instead of eliminating them. Steve Benford, as quoted in  \cite{chalmers2003}, argues that there are four different broad approaches we might take to present seams: (i) a pessimistic approach consists of only showing information that is known to be correct, (ii) optimistic approach is centered around showing all information as if it would be correct, (iii) cautious approach explicitly presents uncertainty and lastly (iv) opportunistic approach seeks to exploit uncertainty. For our conceptual design we determined that the opportunistic approach to seamful design, which in the words of Bill Gaver et al \cite{gaver1992} is \textit{discordant, deliberately leading users to pause or reflect}, would best suit the museum setting. 

While the goals of our project were primarily revolving around the use of BLE technology, the museum context in which this project was carried out prompted us to put some effort into designing a solution that would fit into this environment. The nature of the museum exhibitions thus had some impact on the direction and form of our design solution and as such ought to be mentioned at least briefly. 

First and foremost, the museums were interested in technological solutions that would attract new groups of audiences previously not interested in museum experiences. Young children were identified as one such demographic group that is increasingly out of touch with traditional learning spaces and thus needs more attention. In order to make our solution appealing to this type of audience, two criteria were deemed to be critical; the solution would have to be simple to understand and it would have to be entertaining. Moreover, we were keen to design a solution that would enhance the learning experience traditionally associated with museum visits. To satisfy these requirements, we decided to aim for an educational game-like experience.


Another issue we had to consider was the fact that the Cambridge Museums consortium consists of multiple museum buildings spread across the city. It was thus seen as desirable for our solution to encompass all of these venues. This meant that the game had to be expandable in an organic way to tie together and establish a common user experience across multiple museum buildings. Furthermore the experience had to be designed to take into account that it could potentially be accessed through multiple different entry points (i.e. any one of the individual museums). At the same time we had to take into consideration that it is unlikely for a majority of the visitors to take the time to visit all of the different exhibitions. The game thus also had to be fully functional and accessible by users visiting only one of the museums. 

With these requirements and limitations in mind, we designed an experimental mobile game named \textit{Ghost Detector}. When walking through the museum space with \textit{Ghost Detector} turned on, users would occasionally encounter \textit{ghosts} of various museum artefacts popping up on the screen of the mobile device and challenging the player to find the artefact that they are representing. Studies have suggested that such anthropomorphization of artefacts can produce positive engagement effects, particularly in the case of children.\cite{vazquez2014} 

For this concept to work, every physical artefact associated with a virtual ghost character would have to be equipped with its own BLE beacon. While moving through the museum space, the ghost would then rely on RSS from this beacon to provide users with positive or negative feedback informing them whether they are getting closer or further away. Increases in RSS would thus result in the ghost character getting happy and providing the user with encouraging messages, such as \textit{Yes, I can see we're going into the right direction!} On the other hand situations where the user is moving through an area with weak signal coverage would lead to the ghost character turning angry and reporting to the player that they are getting lost. 

\textcolor{black}{The fact that the strength of the signal produced by each BLE beacon deteriorates over distance was thus used to estimate changes in distance between the user and the artefact. As our preliminary investigation revealed, BLE's signal strength falloff is however not always seamless, making it difficult to provide reasonably accurate distance estimates on a consistent basis. Instead we choose to build a solution that would take advantage of such seams in a way that would contribute to the user experience in a positive way. We did so by encouraging users to treat any feedback from the ghost characters as actual sensory information. For instance, when the user walks into a large crowd, blocking out much of the BLE signal, the ghost would respond with an angry comment such as \textit{I can’t see anything familiar here, I think we’re getting lost!}. The purpose of such a comment is to use obstructed line of sight of the ghost character as a metaphor for sudden drops in RSS. Similarly, in a hypothetical scenario where a user reacts to the negative feedback by raising the smartphone above the crowd, or simply by walking into a more open space, the ghost announces that it can now see where you are going. Although not necessarily correlating with changes in distance from the artefact, such altering feedback thus seems reasonable within the game.} 

In other words by taking into consideration potential problems of the BLE technology and by designing them into the actual game, we were still able to come up with an experience that was expected to function reasonably well despite the dynamic nature of a public museum environment. Figure 2 presents a series of snapshots demonstrating some of the potential feedback a user could encounter. 

\begin{figure}[h!]
 \vspace{-10pt}
\begin{center}
\includegraphics[scale=0.08]{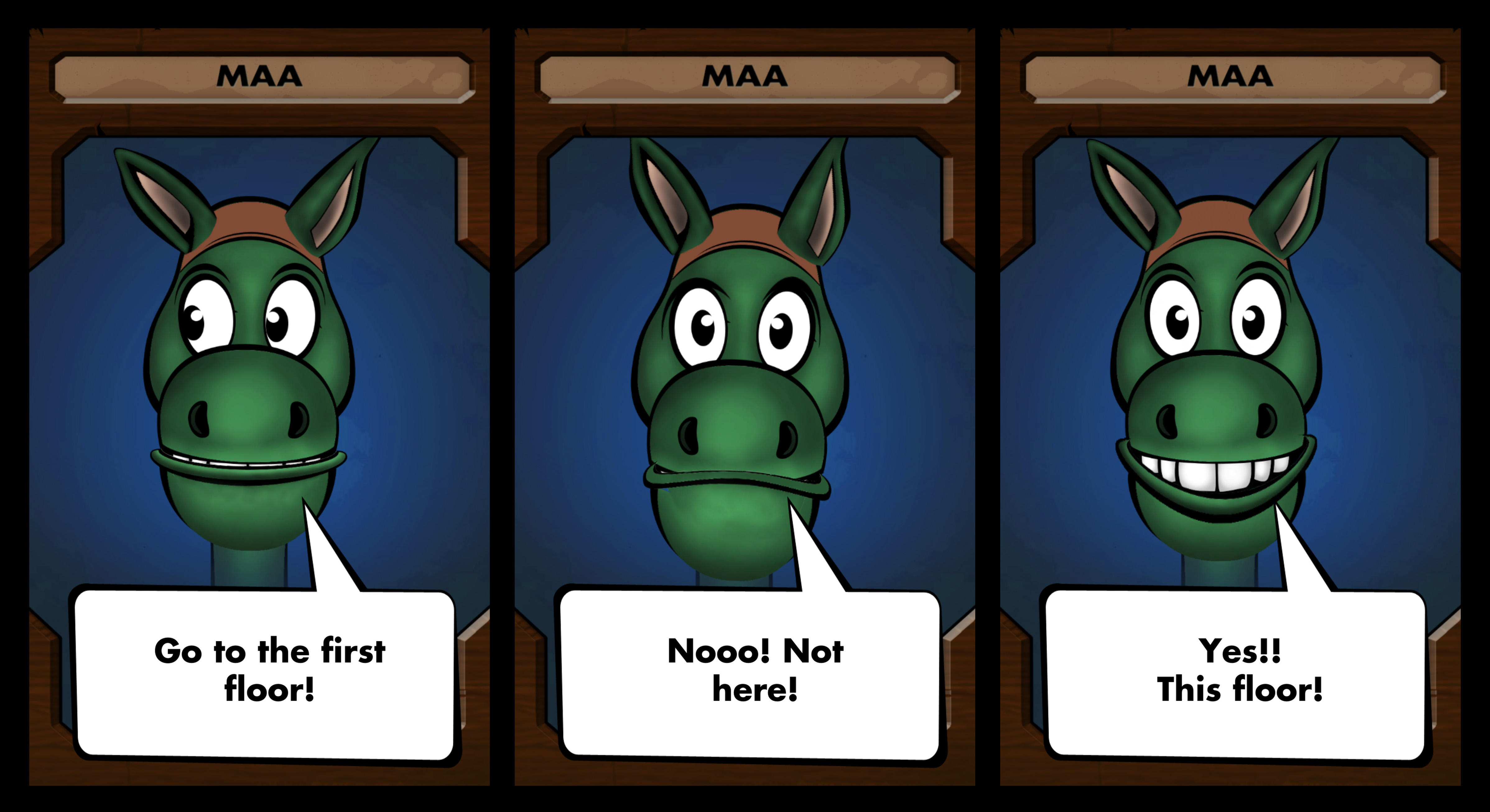}
\caption{Virtual characters representing different exhibits were animated and accompanied by messages to reflect the users' position and situation.}
\end{center}
 \vspace{-15pt}
\end{figure}

Once all of the ghost directed tasks are completed, the user receives an achievement for completing the museum and the option to share it through social media platforms, such as Facebook. Finally, users are contacted by a last lost ghost, which this time explains that it comes from a different museum and needs help finding its way back home. This final ghost serves as an invitation to visit other museum spaces and makes users aware that every museum is a part of a larger experience. Our prototype application is thus essentially a museum guide masking itself as a \textit{Hot and Cold} game. Moreover, as seen in figure 3, our hope was that a game-like application would add a sense of adventure to traditional museum visits, thus resulting in an experience that would be appealing to younger user groups, who are normally perhaps not extremely excited about history or art appreciation. 

\begin{figure}[h!]
\begin{center}
\includegraphics[width=\textwidth]{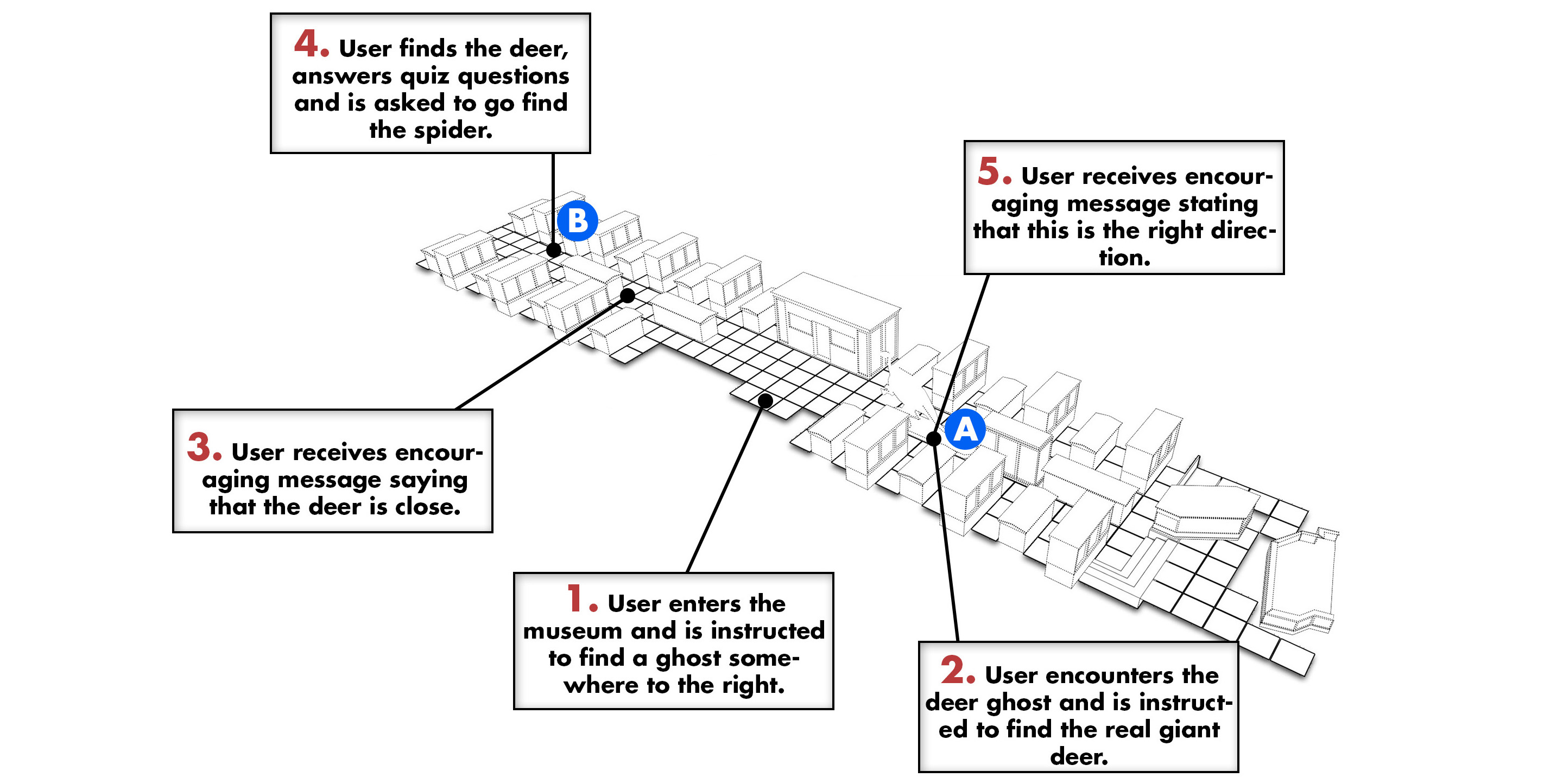}
\caption{Concept illustration depicting the process of helping a ghost deer find its physical counterpart in the University of Cambridge's Sedgwick Museum.}
\end{center}
\end{figure}

\section{Usability Evaluations}
The previous sections of this paper discussed our assessment of the BLE technology and the translation of our findings into a prototype of a potential solution. In order to answer broader questions related not only to the BLE technology but also to its perceived usefulness and social acceptance, as well as the viability of our design approach, we had to evaluate the prototype with real users.

By studying members of the general public and their response to an experience that was presumably radically different from what most of them were used to from traditional museum visits, we were able to gain a better understanding of the strengths, shortcomings and the potentially risky aspects of both a seamful design approach as well as the use of BLE in general. 

In order to gain sufficient insights into the behavior, attitude, emotional response and overall perception of the system, a qualitative phenomenological approach \cite{creswell2013} was deemed to be the most suitable research method. During our first evaluation session we collected data through observations in the form of test monitoring. To give each participant the option to motivate their behavior and explain their impressions, the think-aloud method \cite{tullis2008} was utilized. Once completed, we complemented these observations by conducting interviews with the participants. 
\\

14 groups of users took part in the evaluations. As seen in  Table 1, every major age group was represented in order to provide us with as broad feedback as possible.

\begin{table}[h]
\setlength\extrarowheight{2pt}
\begin{tabular}{  | p{6.0cm} | p{4.0cm}  | c |}
\hline 
User Group & Age Group & Residency \\ \hline 

Mother and daughter & 30s and under 10 & Cambridge \\
A visiting student & 20s & China\\
Museum staff & 30s & Cambridge \\
Grandmother and granddaughter & 60s and under 10 & St Ives \\
Parents with a son and daughter & 40s and under 10 & North England \\
Mother and son & 40s and under 10 & Cambridge \\
Adult female & 50s & Cambridge \\
Grandparents, mother, \& 2 children & 60s, 40s and under 10 & Cambridge\\
Parents and daughter & 30s and 7 & Peterborough\\
Mother and son &  40s and 4 & Cambridge\\
Grandmother, aunt and 3 children & 60s, 40s, 16, 6 and 4 & Cambridge \\
Parents and two daughters & 40s, 14 and 13 & London\\
Parents and daughter & 40s and 7& Cambridge \\
Father and daughter & 30s and 7 & Cambridge \\ \hline

\end{tabular}
\label{users_participating}
\caption{Overview of users participating in the second and third usability evaluation.}

\end{table}

\subsection{Evaluation 1: }
In order to receive some level of feedback as early as possible in the design process, a paper prototype was developed and used during an initial usability evaluation. 
This low fidelity prototype consisted of multiple screen views, each printed on its own A4 paper and each displaying an important event in the game. The first screen view introduced the user to a lost ghost which described the artifact in the museum it wanted to find. Test participants were then asked to move through the museum space to find this given artifact. No further instructions were provided beyond what was stated on the paper prototype screen. 

\textcolor{black}{
Since the project did not yet enter into a stage where a full summative evaluation would be feasible, we determined that a limited number of participants would be sufficient to provide us with a first assessment. A common theory in usability testing, sometimes referred to as the magic number five, claims that about 80 percent of usability related problems will be uncovered by the first five test participants (Lewis, 1994). A similar view has been echoed by Tullis and Albert\cite{tullis2008} who argue that using between 5 to 10 participants is sufficient to accurately assess the major issues. In line with these findings, a total number of 6 participants was recruited for our early evaluation. }
The test subjects were randomly selected out of the museum visitors to roughly represent all major age demographics. Some of them could thus be seen as digital natives, whereas others had a degree of previous exposure to digital technology. We were not familiar with any of the participants before the test session, which meant that no personal relations could affect the answers given in the questionnaires, nor our interpretation of the observations. 

The participants carried out the evaluation one by one and were encouraged to verbalize their thoughts while performing the task. We paid particular attention to any emotional expressions such as frustration or joy. Throughout the evaluation we had one team member walking with the participants and supplying them with new screen views displaying negative or positive feedback. In order to simulate the fluctuating signal strength of BLE beacons, we used the previously developed signal strength maps of the museum space (figure 7) in order to determine what kind of feedback was adequate in various locations of the museum space. 

Once completing the task, test subjects were asked to fill in a questionnaire assessing their impressions of using the prototype. The questionnaire \textcolor{black}{(see figure 7 in the appendix section)} consisted of 5 Likert scale questions, each in the form of a semantic differential spectra ranging between two opposite adjectives (i.e. strongly agree to strongly disagree). The key focus of the questions was to assess the holistic user experience by for instance asking users if they felt the need for additional help while using the prototype or whether they would be willing to use a similar application more frequently. Additionally, 4 open-ended questions were included in the questionnaire, giving the participants the opportunity to express themselves more specifically regarding their likes and dislikes about the system. These were later analyzed using the affinity diagramming method \cite{tullis2008} in an attempt to find any reappearing theme. 

\textcolor{black}{
The data collected through these questionnaires in combination with our observations  resulted in a number of relevant findings: \begin{itemize}
	\item In spite of interacting with a paper prototype, participants had no problem approaching it as a real mobile device while carrying it with them throughout the museum. This might be a good indication of the widespread social acceptance of mobile devices. In fact, it appears that they are becoming so deeply rooted in our culture that we now approach them naturally without any hesitation. The only aspect of the system which participants described as being novel to them was the location awareness. Nonetheless, they were all able to pick up the idea rather quickly. Multiple users even spontaneously began suggesting innovations to the application, such as adding characters representing additional museum artefacts that would be challenging to find. 
\item There was however one case where a participant experienced minor problems in completing the task. One child of preschool age began walking in the right direction and accordingly received positive feedback. In that moment he however turned around and began walking in the opposite direction. This triggered negative feedback, which made him pause and reflect the situation for a brief moment. Eventually though he managed to figure out the correct direction and find the artefact. The fact that even a small child managed to complete the task, albeit after a brief struggle, provides an indication that the concept is child-friendly. 
\item Once locating the target artifact, children typically manifested genuine joy at succeeding to complete the task. This degree of positive reaction was not observed among any of the adult participants. In fact many of them openly admitted that they found the experience to be geared towards children. 
\item The questionnaires largely confirmed the observations made during the test monitoring, with most of the answers being positive towards the application. \textcolor{black}{As seen in figure 4, none of the respondents gave a negative reply to any of the likert scale questions.} \textcolor{black}{All of the participants agreed that they did not need any additional help while using the application.} All of the participants also stated that they either \textit{strongly agree} or \textit{agree} that the application enriched their museum experience. Furthermore, many participants stated that they found the application innovative, a claim which was reinforced by the fact that all of the participants answered that they've never encountered any similar application before. 
\end{itemize}
} 
\begin{figure}[h!]
 \vspace{-15pt}
\begin{center}
\begin{center}
\includegraphics[width=\textwidth]{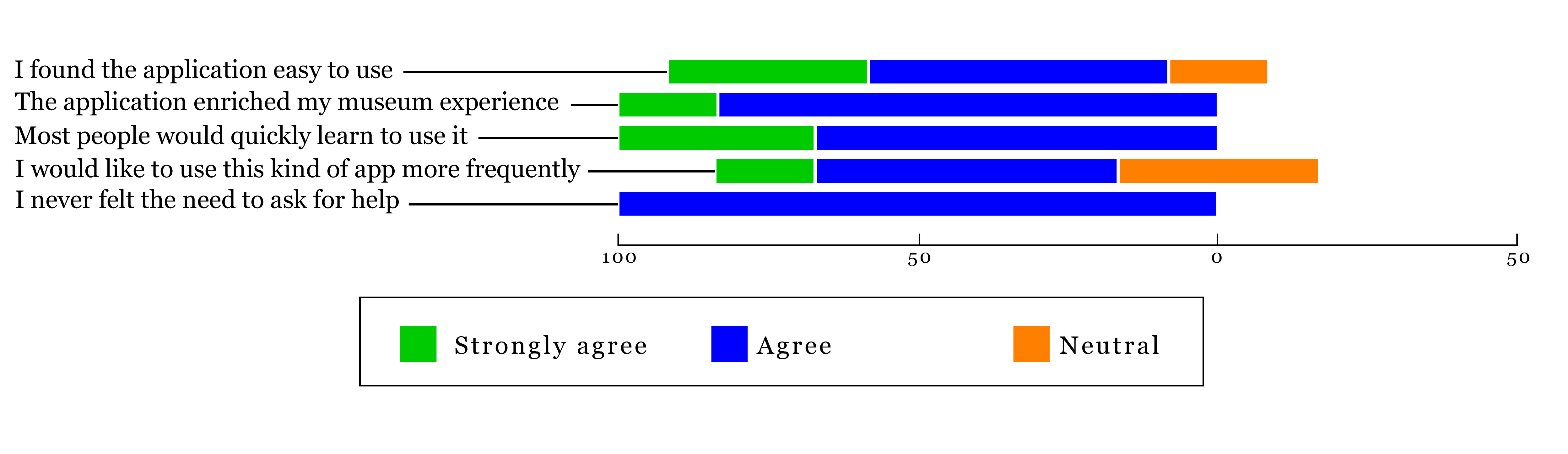}
\end{center}
\vspace{-18pt}\textcolor{black}{
\caption{The first evaluation, carried out with a paper prototype, generated predominantly favourable responses from the participants.}}
\end{center}
 \vspace{-15pt}
\end{figure}

\subsection{Evaluation 2: }
Once we had real data indicating the viability of our concept, the next step was to develop and evaluate a high fidelity prototype. A second usability evaluation was thus conducted with an actual working prototype of our application and a set of BLE beacons deployed in one of the University museums. Our key goal with this second study was to assess ways in which the core experience of visiting a museum would be impacted by introducing our BLE-enabled application as well as any behavioral changes manifested by the application users compared to regular visitors. 

Much like in the first evaluation, museum visitors were invited to use the application to locate a specific museum exhibit. Six sets of participants took part in this evaluation; four young families and two individual adults. Once consent to take part in the study was obtained, participants were handed an Android phone running the application and were instructed to start walking through the exhibition. The prototype application then provided users with feedback through visual as well as audio elements. Abstract negative or positive sounds (e.g. happy laughter or an angry scream) were used to attract participants' attention to the application, causing them to look at it had they been looking elsewhere. They would then see a corresponding animation of the ghost character accompanied by a speech bubble with a brief text message containing appropriate feedback. Two researchers observed from distance how participants interacted with the application and with the museum environment. Once done with the task, every group of participants took part in a semi-structured interview. The focus of this interview was to gain more detailed understanding of the general thoughts users had about the use of our location aware application. 

\textcolor{black}{
The potentially important information surfacing as a result of this evaluation can be summed up as follows:
\begin{itemize}
\item Throughout the evaluation participants were found to act in a very goal directed manner when using the application, moving purposefully through the museum towards their goal, rather than browsing the exhibits. 
\item In the case of multi-user groups, such as families, children would typically \textit{race off} with the phone and try to find the exhibit while parents would follow.
\item Participants were able to clearly understand the feedback provided to them by the application and were able to tell whether they were moving closer or further away from the artefact they were asked to find. Animated facial expressions of the ghost characters in combination with sounds proved to be effective in guiding. The majority of participants did however report relying most heavily on text messages to lead them to the exhibit. 
\item When interviewed, participants were generally positive to the application. Multiple users liked the playful way in which the application led them to an exhibit, with children expressing lots of excitement and enthusiasm when using it and parents voicing their approval of an interactive educational tool for their children. 
\item We did however come across a number of negatives as well. Aside from a few minor GUI related complaints, participants expressed reservations about the sounds produced by the application. They were concerned that sounds would be disruptive to other museum-goers, especially if a large number used the application at once. The sounds also caused them to feel somewhat self-conscious whilst using the application. This could naturally have been fixed simply by turning off the sound, however that proved to lead to another, potentially even more serious issue. By turning off the sound, users had to rely entirely on visual cues displayed on the smartphone screen. While searching for the given exhibit with the sound turned off, users kept wandering through the museum space with their eyes glued to the smartphone screen, rather than enjoying the exhibition. One participant explicitly mentioned that the attention they had to pay to the application damaged their museum experience. Another participant, who accompanied her young granddaughter while she used the app, felt that she was too engrossed in the application, and frequently had to draw her attention back to the museum artefacts she was missing. This was clearly in conflict with our previously stated goal of developing an application that would complement the museum space, rather than compete with it.
\end{itemize}
}
 
\subsection{Evaluation 3: }
The outcome of the second evaluation made us abandon the idea of providing visitors with feedback in real time. Instead we implemented a popup message that appeared whenever new feedback was available for the user to view. This message was accompanied by a subtle vibration signal which was repeatedly played until the user triggered the feedback by tapping on the popup message. Our intention was that such a system would allow visitors to pay attention to the museum space instead of constantly having to watch the smartphone screen.
As another notable adjustment, we decided to scale up the whole system to involve multiple exhibits across three of the University museums. 

\textcolor{black}{This in itself introduced another challenge, namely the fact that one of the newly introduced museums was hosting an exhibition taking place across two floors, making BLE positioning increasingly delicate. In order to prevent the signal received from beacons located at one floor from interfering with signals coming from beacons at another floor, every stairway connecting the different floors had to be equipped with two beacons; one at the bottom and one at the top of the stairway. These stairway beacons functioned as \textit{switches}, prompting the application to ignore any signal detected from beacons whose ID was associated with a different floor. For instance, when the user moved into a low proximity of the top stairway beacon, the application knew the user was located at the top floor and thus stopped taking into consideration all signals coming from beacons located at the ground floor. Similarly, when a user took the stairway back to the ground floor, crossing the RSS threshold of the bottom stairway beacon would automatically reverse the application into only acting upon signal pulses from beacons located at the ground floor.}

Additionally, in order to increase the educational value, we implemented a set of quiz questions that users were challenged to answer after successfully completing any of the ghost-directed tasks. 
Besides assessing these new features, our third study was once again designed to evaluate how our BLE-enabled application impacted users’ experience. This time, we did however pay a particular attention on the implications of having the experience stretched to encompass multiple exhibits and locations.  

A total number of eight sets of participants took part in this study. Two were pre-recruited, while the remaining 6 were recruited on the day at the museums. Test subjects were invited to use the application to locate 6 exhibits across 3 museums. Once consent to take part in the study was obtained, participants were handed an Android phone running the application and asked to follow instruction leading them to the exhibits. As with previous evaluations, no further instructions were given. The researcher observed how participants interacted with the application and their environment as they completed the task. After the evaluation, participants were asked a series of questions about their experiences using the application. 

\textcolor{black}{
A number of relevant observations was made throughout this evaluation session:
\begin{itemize}
\item The new popup message system increased users' immersion with the museum environment. In contrast to the previous version of the application prototype, people appeared to spend less time looking at the screen. Users typically glanced at the screen between messages just to see if the instructions had changed. They did however quickly learn that new instructions only appear as messages accompanied by vibration. 
\item The game proved to work well when played by groups of people, such as families. In some cases the phone was held by an adult who would encourage children to trigger newly received feedback messages. The adult then read the speech bubble aloud and asked children which way they thought the group should proceed. In other cases the phone was held by a child who spearheaded the search for an exhibit whilst a parent followed closely behind to give help or explanation if needed.
\item Answering quiz questions appeared to foster a deeper engagement with the exhibits, as people spent time reading information they would otherwise have ignored. While groups would sometimes fragment when looking for exhibits, they tended to reconvene to complete the quiz, which proved the more collaborative of the two activities. Both children and parents reported that answering the questions encouraged them to explore exhibits more thoroughly. Parents felt the quiz gave their children something to focus on, instead of just briefly looking at the exhibit before moving rapidly onto the next. 
\item Scaling up the experience across multiple museums introduced some additional challenges. Once completing all the ghost directed quests in one museum, users received a message encouraging them to cross over to another museum. By leaving the museum space, users however also left the environment covered by BLE beacons and entered dead space, where no feedback or guidance could be given by the application. Even though all the museums taking part in this evaluation were within walking distance of each others, some users still reported that they found the lack of guidance to be a problem and suggested that the app ought to provide them with a map to the next location.
\end{itemize}
}

\begin{figure}[h!]
\begin{center}
\begin{center}
\includegraphics[scale=0.06]{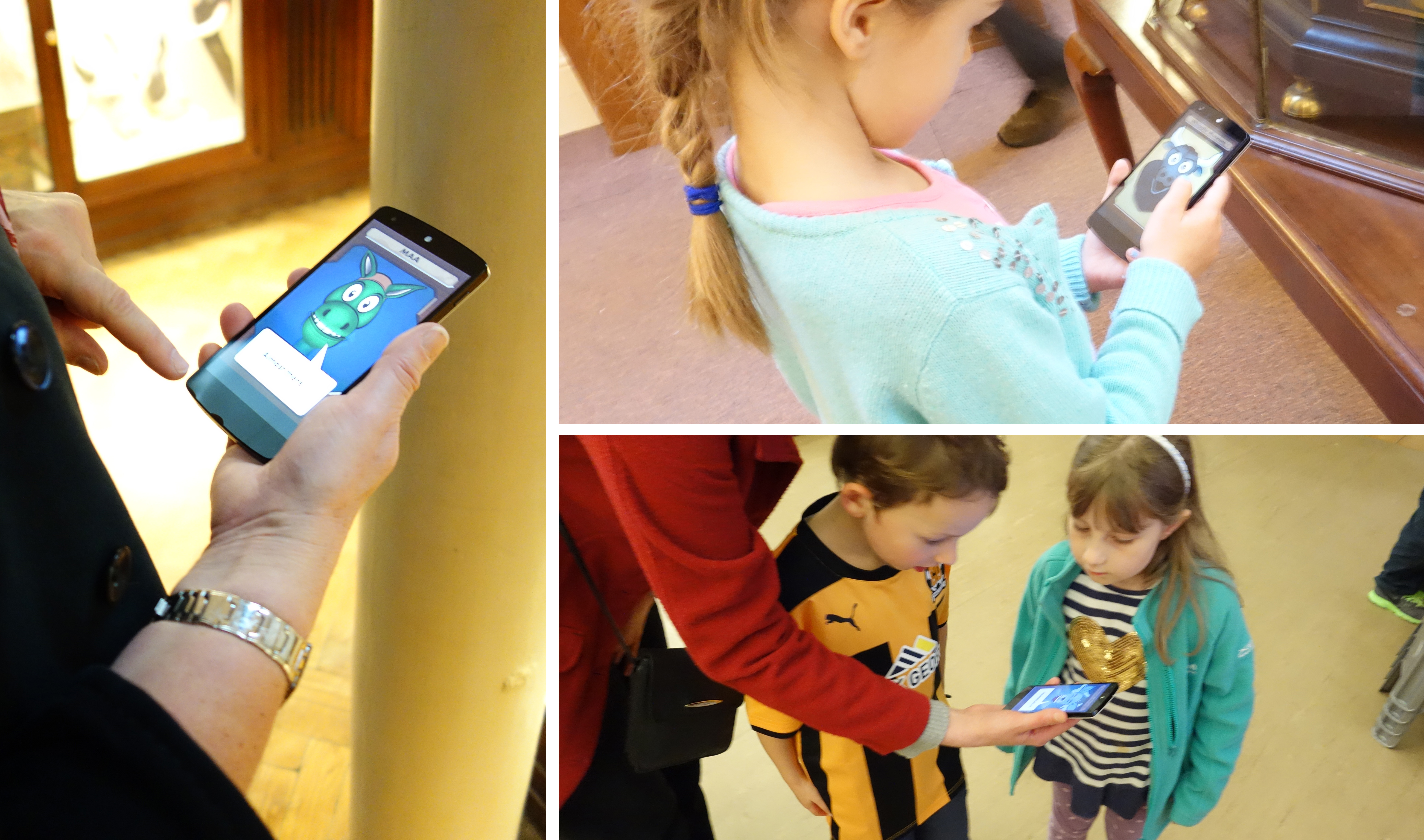}
\end{center}
\caption{A number of families, including small children, took part in the evaluation process. Image copyright: Dovetailed}
\end{center}
\vspace{-15pt}
\end{figure}
\subsection{Comparison to Other Seamful Design Experiments}
\textcolor{black}{As described in the \textit{related work} section of this paper, a number of other seamful location-based games has been deployed and evaluated in the past, most notably CYSMN and Bill \cite{benford2002}. Table 2 highlights some of the similarities and differences between these projects and the Ghost Detector}
 


\begin{table}[h]
\begin{center}
\setlength\extrarowheight{5pt}
\begin{tabular}{ | p{3.0cm} | p{3.0cm}  | p{3.0cm} | p{3.0cm} |}
\hline 
\textbf {Game} & \textbf{Can You See Me Now?} & \textbf{Bill} & \textbf{Ghost Detector} \\ \hline \hline 

\textbf{Year of deployment} & 2001 & 2004 & 2014 \\ \hline

\textbf{Evaluation participants} & 214 players over the internet, 3 runners physically present & 56 & 20 groups featuring a total number of 36 participants\\ \hline

\textbf{Number of evaluation sessions} & 2 initial, several more staged later for demonstration purposes & 3 & 3 (1 featuring a paper prototype, 2 featuring a working application)\\ \hline

\textbf{Duration of the evaluation} & 6.5 hours split over 2 days & 4 hours split over 3 days & Approximately 12 hours split over 3 days\\ \hline

\textbf{Duration of the user experience} & Ranging from 13 seconds to 50 minutes & 15-25 minutes & Ranging from 5 minutes to 1 hour\\ \hline 

\textbf{Distance traveled} & varies & varies & approximately 300 meters \\ \hline 

\textbf{Types of collected data} & 3 (audience feedback, ethnographic studies and analysis of system logs) & 3 (Video recordings, observations and game logs) & 3 (Questionnaires, observations and interviews) \\ \hline 

\textbf{Cost of hardware} & 
\pounds £149.99 (Compaq iPAQ Pocket) & 
\pounds£212.99 (HP iPAQ H5550 Pocket PC) & \pounds£85.00 (Samsung Galaxy S3 Smartphone) \\ \hline 

\textbf{Positioning accuracy} & 
~3.5 meters (GPS) & ~3.5 meters (GPS) & ~2.6 meters (BLE) \\ \hline 

\end{tabular}

\label{users_participating}
\textcolor{black}{
\caption{A summary of the evaluation of Ghost Detector and two other notable seamful games}
}
\end{center}
\end{table}

\section{Discussion}
The purpose of the research brought forth in this paper was to assess the viability of BLE as an enabling technology for location-based mobile experiences and to identify a suitable design approach that would allow us to use its potential to improve user experience in public spaces. 

The novelty of BLE meant that there was not much ground to build on in terms of previous research. One of the first extensive studies conducted on this topic was the aforementioned survey by Faragher and Harle \cite{Faragher2014b} which provided a clear indication that the precision of BLE is currently not sufficient for enabling sophisticated context-aware tasks requiring accurate positioning. The results produced by our preliminary investigation clearly supported these findings. 

Nonetheless, this per se does not automatically mean that BLE cannot at its present state be used to build innovative location-based experiences in spaces of our daily lives. We argued that a seamful design framework, as proposed by Broll and Benford \cite{benford2005}, offeres a way of overcoming some of the problems we encountered. Oulasvirta et al. \cite{Oulasvirta2004} argued that the process of seamful design consists of solving three key problems; (i) Understanding of which seams are important, (ii) presenting the seams to the users and finally (iii) designing interactions with these seams. Following this reasoning, the most obvious seams we had to deal with in our project were constituted by signal strength falloff over distance and by unpredictable fluctuations arising as a consequence of spatial factors beyond our control (e.g. uneven and complex museum landscape or people randomly walking around and interfering with the signal propagation). In the proposed solution we thus attempted to present these seams by translating signal strength falloff into a distance estimate, with greater RSS values being interpreted as the user being closer to a BLE beacon, while weaker RSS meant the user was further away. Although objects in the environment frequently hampered the BLE signal propagation and thus distorted this distance estimate, we attempted to make these inconsistencies look like a natural part of the game by for instance pitching distance estimate as the field of view of a ghost character. 

Through a set of usability evaluations, we were able to verify the feasibility of this concept. The participants in our user centered evaluation appeared to appreciate the Ghost Detector by claiming that it enriched their experience and that they would like to use similar applications more frequently. In fact the evaluation outcomes are indicating a high level of social acceptance of this type of experiences. Moreover, our study suggested that by taking this approach, the user experience will be particularly improved for children. Since attracting new demographic groups was one of the main design goals of our project, it is feasible to sum up our work as a successful implementation of BLE-enabled experience using seamful design approach. Satisfactory results were thus reached without having to adopt the strive for seamlessness that so often dominates contemporary design. 

It is however imperative to keep in mind that a range of comparable experiences has been produced in the past using a traditional seamless design approach. As mentioned in the \textit{related work} chapter, a similar system has already been successfully designed and implemented into a Louvre exhibition \cite{Arnaudov2008}. We cannot therefore claim that a seamful design framework is the only valid approach when developing mobile location-based experiences. We do however feel inclined to maintain that the high complexity and unpredictability of most public environments means that designers of all sufficiently advanced systems will have to deal with the fact that no technical solution can be expected to operate perfectly consistently. Particularly so in the case of a technology as versatile and ambitious as the BLE. \textcolor{black}{A more detailed comparison between these two paradigms is presented in the following table III. }

\begin{table}[h]
\setlength\extrarowheight{5pt}
\begin{center}
\begin{tabular}{|p{4.2cm} | p{4.2cm}  | p{4.2cm} |}
\hline 
\textbf{Design approach} & \textbf{Seamless} & \textbf{Seamful} \\ \hline \hline 

\textbf{Conceptual metaphor} & World as a map to be read & World as a puzzle to be discovered \\ \hline

\textbf{Calibration} & Comprehensive survey is used to register fingerprint regions with respect to ground truth grid measure & Partial survey is used to determine threshold boundaries with respect to particular locations \\ \hline 

\textbf{Coordinate system} & Absolute positioning on coordinate grid &Partial ordering of semantically distinguished locations
 \\ \hline 
 
\textbf{Multipath accommodation} & System must recognise and disambiguate locations that share reflected signal mappings & Location ambiguity is used as an explicit feature in the application design \\ \hline 
 
\textbf{Accuracy assumption} & Uniform tolerance margin throughout sensing space & Variable tolerance margin at different positions\\ \hline 

\textbf{Game content delivery} & Focus on content being visible at some locations & Focus on content being invisible at some locations \\ \hline 

\textbf{Navigation behaviour} & Heads-down, with continuous update & Heads-up, with intermittent update
\\ \hline 

\textbf{Required user behaviour} & Users must take care not to obstruct signal or shield antenna &  Users are encouraged to move in response to variability of location data \\ \hline 

\end{tabular}
\label{seamless_seamful}
\textcolor{black}{
\caption{Overview of  key differences between a seamless and seamful design approach}
}
\end{center}
\end{table}


Although both the human centered evaluation and the testing of the BLE technology was carried out in the museum setting, we do not believe that our results should be looked upon as context-specific and non generalizable. Many of the technology-related limitations of BLE encountered in the museum context are likely to be present in other public environments as well. Due to the common nature of visitors, there is no reason to believe that the attitudes manifested by users towards the technology in the museum space would not correlate with those we would encounter in other public environments. Consequently, it seems reasonable to believe that the potential solutions brought forth in this paper could be applicable in a broad spectrum of settings.

However, the results presented in our study ought to also be taken with a degree of caution. Particularly so because the scope and ambition of the usability tests were rather limited. The  relatively small sample of users participating in our evaluations did naturally not pose a stress test for the concept and in turn cannot warrant any conclusive statement regarding the future potential of this design approach. Rather, our work ought to be regarded as a case study centered around an experiment carried out in a constrained time and space. Our ambition in this sense can thus be described as putting a spotlight on the potential benefits of unconventional design approaches when developing novel ubicomp solutions.

\textcolor{black}{
\subsection{Design recommendations}
A number of design recommendations can be extracted from the results of our study: \begin{itemize}
  \item In the context of a cultural space, such as a museum, implementation of new digital technology may end up distracting users from the original content. As our study demonstrated, information presented and updated in real time will inevitably compete for users’ attention. A notification system using a subtle signal, such as a vibration tone, to let the user know whenever new digital content is available, proved to be less disruptive to the original experience. 
\item Persuading users to use novel technology (or a familiar technology in a novel context) represents a challenge. Based on the technology acceptance theory, we argued that as long as potential users perceive the technology as being easy to use and as long as they are able to see some form of benefit associated with its use, there is a substantial probability that the technology will reach a meaningful level of social acceptance. We do therefore believe that successfully communicating these two values to potential users ought to be seen as a priority for future designers working with digital experiences in public spaces. 
\item The range of BLE beacons is naturally limited. When designing location aware applications involving artifacts distributed across larger distances, such as for multiple exhibits located in different museums, developing a hybrid positioning system involving GPS might be a good way to bridge the space beyond reach of BLE beacons. 
  \item Embracing RSS inconsistencies and turning them into a part of an experience is a viable approach for design of ubicomp systems, particularly so when it comes to gamifying a cultural space. 
  \item Unpredictable RSS fluctuations and a consequently inaccurate positioning can be partially accommodated through the use of metaphors, such as pitching RSS as the field of view of a virtual character. 
\end{itemize}
}
\vspace{-8pt}

\section{Conclusions} 
\textcolor{black}{More than two decades have now passed since Weiser made his prediction of a world in which people would be constantly immersed in digital information. \cite{weiser1993} Since then, a range of solutions has been successfully adopted to augment the physical spaces of our daily lives with a layer of digital information. Thanks to its versatility, low power consumption and comparatively high data transfer rate, BLE represents arguably the most ambitious step towards realizing Weiser's vision.} 

\textcolor{black}{In this paper we highlighted some of the design challenges associated with the development of BLE-enabled ubiquitous information solutions for public environments. \textcolor{black}{By revealing and exploiting the infrastructure problems (or \textit{seams}, as we have come to call them) that were originally hampering the functionality of this technology, we were able to demonstrate the viability of a seamful design framework. Moreover, we have argued that given the increasing ubiquity of location-based information services out in the wild, this design approach is likely to see a growth in popularity in the near future. Finally, our study suggests that a seamful museum experience has the potential to attract new audiences, such as children, who are typically not interested in cultural or educational content. }}


\textcolor{black}{Although our aim was to develop a system that would merely complement the original experience (i.e. a museum visit), without rivalling it, we did observe a certain degree of behavioral change among the users participating in our evaluations. Children racing through museum corridors while paying more attention to the feedback on their smartphone than the actual museum exhibits around them, provided a clear indication that their way of experiencing the museum exhibition was indeed affected by the introduction of a ubiquitous information system.}

\textcolor{black}{It is arguably inevitable that any major technological innovation will alter the way we go about our daily life to some extent. The much quoted media philosopher Marshall McLuhan remarked that “\textit{we become what we behold. We shape our tools and then our tools shape us}” \cite{mcluhan1964}. The truth behind this statement has perhaps never been more apparent than it is today. And in many ways the idea of a seamful design approach appears to resonate with such a technologically deterministic view. By arguing for an embracement of the inherent infrastructure problems instead of attempting to fix their underlying causes, one could easily raise the objection that our approach seeks to adapt user experiences to the available technological solutions, rather than the other way around. In other words, it might seem as if we are willing to give in to technology and allow it to shape our behavior.}

\textcolor{black}{However, what we have been arguing for is in fact merely a shift of perspective; a refocus from the engineering practices producing a technology, to the design frameworks responsible for its embedding into a social context. Seamful design teaches us that, from an end-user point of view, what really matters is not as much the technology itself, but rather the way we choose to use it. And perhaps in this lies the most significant moral surfacing as a result of the work described in this paper. McLuhan had every right to stress the profound impact technology has on our lives,  however we ought to never succumb to the feeling that we are at the mercy of technological progress. Although every technology does come with a certain set of limitations and predispositions, these are never unalterable. A sensible design approach is a powerful tool through which we can convert technological weaknesses into strengths, implement them into our systems, even turn them into their driving elements, and ultimately pave the way for a better user experience.} 

\begin{acks}
The authors would like to thank the University of Cambridge Museums consortium, particularly the Sedgwick Museum, the Museum of Anthropology and Archaeology and the Whipple Museum, for granting us full access to their museum displays. Cambridge Silicon Radio provided the necessary hardware. We would also like to thank Rob Harle and Ramsey Faragher for their technical assistance. Dovetailed, Cambridge, assisted with the final phase of user evaluation. 
\end{acks}

\hfill

\vspace{-12pt}
\bibliography{bib2}
\bibliographystyle{apalike}

\hfill
\hfill
\appendix
\section*{APPENDIX}
\setcounter{section}{1}

\appendixhead{ZHOU}

\begin{figure}[h!]
\begin{center}
\begin{center}
\includegraphics[width=\textwidth]{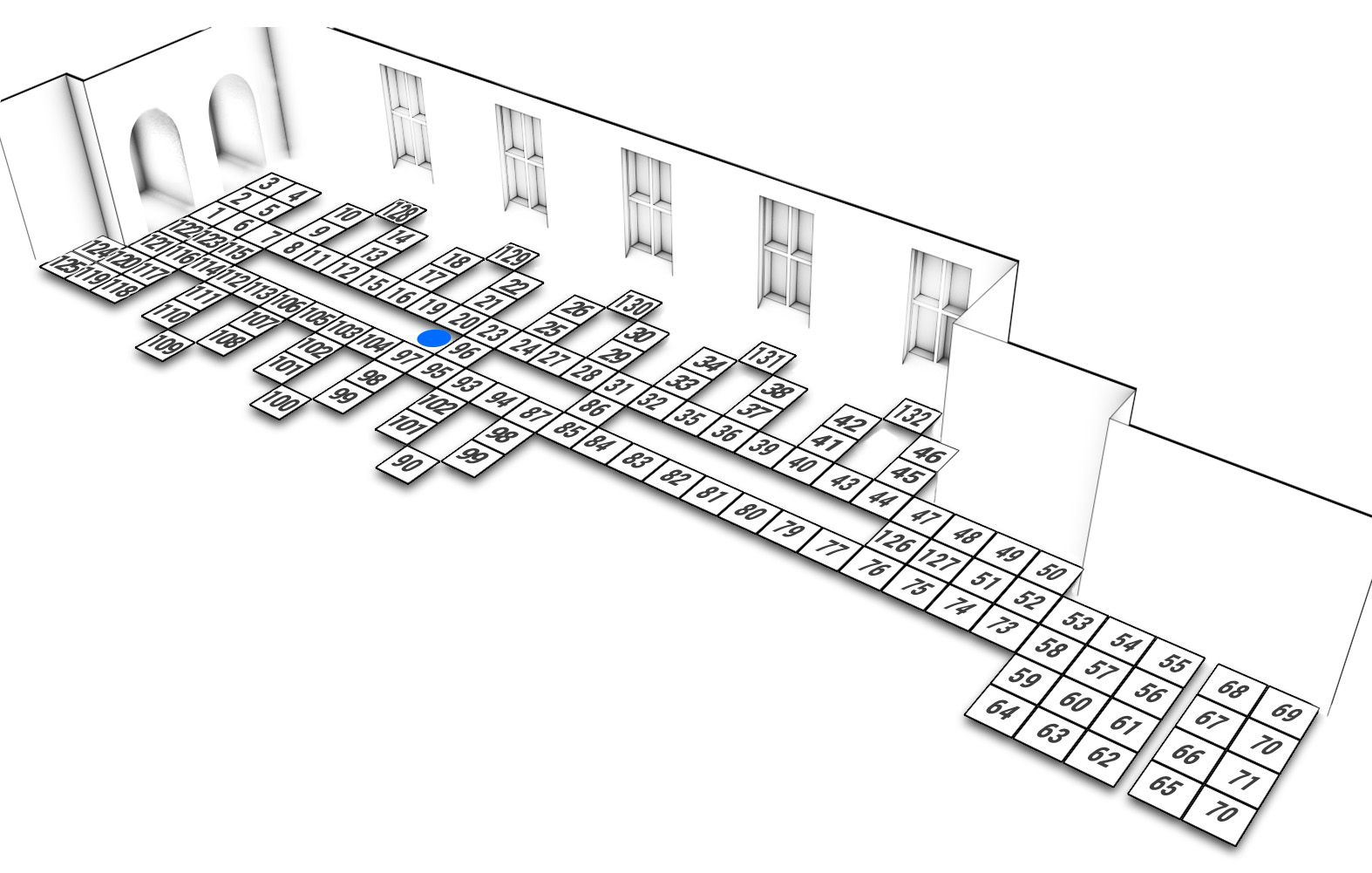}
\end{center}
\caption{A map depicting RSS sampling points in the East Wing of Sedgwick Museum of Earth Sciences. Every numbered square represents 1$m\textsuperscript{2}$. The blue dot represents location of the beacon. Specific RSS values for each sampling point are depicted in table IV and V.}
\end{center}
\end{figure}

\begin{table}[htbp]
\footnotesize 
\centering
\setlength\extrarowheight{1pt}
\begin{tabular}{|M|M|M|M|M|M|M|M|M|}
\hline
\multicolumn{ 9}{|c|}{\textbf{East Wing - Beacon1}} \\ \hline
\multicolumn{ 1}{|c|}{\textbf{Location}} & \multicolumn{2}{c|}{\textbf{South}} & \multicolumn{ 2}{c|}{\textbf{East}} & \multicolumn{ 2}{c|}{\textbf{North}} & \multicolumn{ 2}{c|}{\textbf{West}} \\ \cline{ 2- 9}
\multicolumn{ 1}{|l|}{} & \textbf{RSS} & \textbf{SD} & \textbf{RSS} & \textbf{SD} & \textbf{RSS} & \textbf{SD} & \textbf{RSS} & \textbf{SD} \\ \hline\hline
1 & -81 & 5,5 & -80 & 1,7 & -88 & 2 & -85 & 2,8 \\ \hline
2 & -84 & 3,6 & -84 & 2,7 & -86 & 2,7 & -86 & 4 \\ \hline
3 & -89 & 2,2 & -88 & 3,2 & -83 & 1,7 & -80 & 2,8 \\ \hline
4 & -86 & 3,7 & -88 & 1,4 & -85 & 1,4 & -82 & 1,4 \\ \hline
5 & -88 & 3,3 & -84 & 2,2 & -89 & 1 & -87 & 2 \\ \hline
6 & -85 & 3,2 & -79 & 1,4 & -81 & 1,4 & -85 & 3,6 \\ \hline
7 & -79 & 5,5 & -82 & 3,2 & -82 & 2,7 & -88 & 2,2 \\ \hline
8 & -80 & 5,1 & -82 & 3,2 & -80 & 2,7 & -85 & 3,3 \\ \hline
9 & -78 & 4,3 & -79 & 5,1 & -77 & 2,5 & -87 & 2,7 \\ \hline
10 & -82 & 3,5 & -80 & 3,9 & -84 & 3,9 & -85 & 3,5 \\ \hline
11 & -75 & 2,7 & -75 & 3 & -81 & 5,4 & -85 & 4 \\ \hline
12 & -82 & 3 & -75 & 4,2 & -79 & 3,5 & -81 & 4,1 \\ \hline
13 & -82 & 1 & -82 & 2 & -80 & 6,2 & -86 & 3,3 \\ \hline
14 & -82 & 4 & -80 & 2 & -84 & 3,6 & -83 & 2 \\ \hline
15 & -78 & 5,4 & -77 & 3,3 & -73 & 4,4 & -81 & 5,3 \\ \hline
16 & -74 & 5,8 & -69 & 3,2 & -66 & 2,2 & -81 & 5,9 \\ \hline
17 & -76 & 4,4 & -71 & 3,6 & -80 & 3,9 & -75 & 4,4 \\ \hline
18 & -72 & 2,5 & -76 & 3,3 & -80 & 3,5 & -80 & 4 \\ \hline
19 & -71 & 4,6 & -63 & 1,4 & -68 & 5,7 & -75 & 3,4 \\ \hline
20 & -76 & 3,9 & -70 & 1 & -80 & 5,6 & -77 & 4 \\ \hline
21 & -72 & 3,3 & -73 & 5,1 & -82 & 3,3 & -77 & 4,4 \\ \hline
22 & -73 & 2,8 & -76 & 1,4 & -84 & 4,4 & -79 & 2,7 \\ \hline
23 & -77 & 2,8 & -75 & 2,5 & -79 & 3,2 & -76 & 2,8 \\ \hline
24 & -75 & 3,3 & -81 & 2,7 & -83 & 3,3 & -79 & 2,8 \\ \hline
25 & -80 & 5,1 & -85 & 2 & -85 & 4,8 & -79 & 3,5 \\ \hline
26 & -79 & 2 & -85 & 3,7 & -86 & 3,3 & -80 & 2,8 \\ \hline
27 & -76 & 3,7 & -82 & 6 & -81 & 3,2 & -81 & 2,8 \\ \hline
28 & -77 & 4,2 & -84 & 2 & -78 & 2 & -83 & 3,9 \\ \hline
29 & -84 & 3,3 & -84 & 3,6 & -84 & 3,6 & -81 & 2,8 \\ \hline
30 & -83 & 2,8 & -83 & 2,8 & -84 & 3,2 & -85 & 2,5 \\ \hline
31 & -80 & 4 & -90 & 1 & -79 & 3,6 & -79 & 3,2 \\ \hline
32 & -82 & 4,6 & -86 & 5,2 & -81 & 4,5 & -78 & 2 \\ \hline
33 & -85 & 3 & -88 & 2,8 & -89 & 2 & -91 & 1 \\ \hline
34 & -86 & 2,7 & -89 & 2,2 & -87 & 2,7 & -89 & 2 \\ \hline
35 & -85 & 2,7 & -86 & 6,7 & -81 & 3,2 & -84 & 3,7 \\ \hline
36 & -80 & 3,6 & -86 & 2,1 & -86 & 2,7 & -80 & 2,8 \\ \hline
37 & -88 & 2,7 & -89 & 1,4 & -87 & 2,7 & -86 & 3,6 \\ \hline
38 & -86 & 3,3 & -87 & 3,5 & -84 & 1 & -88 & 2 \\ \hline
39 & -88 & 3,7 & -89 & 2,7 & -87 & 2,2 & -84 & 2,7 \\ \hline
40 & -84 & 3 & - & - & -88 & 2 & -84 & 2 \\ \hline
41 & -89 & 3,5 & -90 & 1 & - & - & -89 & 2 \\ \hline
42 & -82 & 2 & -82 & 2,5 & -81 & 4 & -82 & 2,5 \\ \hline
43 & -78 & 4,1 & -77 & 4,4 & -78 & 4,6 & -79 & 4,5 \\ \hline
44 & -81 & 3,2 & -86 & 3,3 & -85 & 3,2 & -83 & 3,7 \\ \hline
45-79 & - & - & - & - & - & - & - & - \\ \hline
80 & -89 & 2,2 & - & - & -86 & 2,2 & -87 & 1,7 \\ \hline
\end{tabular}
\caption{RSS values fluctuated heavily based on the orientation of the user. Sampling points referred to in the table are depicted in figure 6.}
\end{table}

\begin{table}[htbp]
\footnotesize
\centering
\setlength\extrarowheight{1pt}
\begin{tabular}{|M|M|M|M|M|M|M|M|M|}
\hline
\multicolumn{ 9}{|c|}{\textbf{East Wing - Beacon1}} \\ \hline
\multicolumn{ 1}{|c|}{\textbf{Location}} & \multicolumn{ 2}{c|}{\textbf{South}} & \multicolumn{ 2}{c|}{\textbf{East}} & \multicolumn{ 2}{c|}{\textbf{North}} & \multicolumn{ 2}{c|}{\textbf{West}} \\ \cline{ 2- 9}
\multicolumn{ 1}{|l|}{} & \textbf{RSS} & \textbf{SD} & \textbf{RSS} & \textbf{SD} & \textbf{RSS} & \textbf{SD} & \textbf{RSS} & \textbf{SD} \\ \hline\hline
81 & -88 & 2,7 & - & - & -86 & 3,3 & -87 & 1,7 \\ \hline
82 & -87 & 3,5 & - & - & -84 & 3,5 & -83 & 2,8 \\ \hline
83 & -83 & 4,9 & -87 & 3 & -82 & 4,6 & -87 & 2,7 \\ \hline
84 & -86 & 2,7 & -90 & 1,7 & -84 & 2,7 & -86 & 3 \\ \hline
85 & -86 & 3 & -89 & 1,4 & -86 & 4,5 & -84 & 3,9 \\ \hline
86 & -85 & 3,1 & -88 & 2,8 & -84 & 3,5 & -83 & 3,2 \\ \hline
87 & -84 & 3,5 & -86 & 5,3 & -85 & 4,9 & -83 & 3,1 \\ \hline
88 & -86 & 3 & -83 & 3 & -81 & 4,1 & -84 & 3,9 \\ \hline
89 & -84 & 2 & -83 & 2,2 & -84 & 3,3 & -86 & 2,7 \\ \hline
90 & -86 & 3 & -83 & 4 & -80 & 2,8 & -85 & 4,2 \\ \hline
91 & -89 & 1,9 & -85 & 3,7 & -85 & 4,1 & -86 & 3,2 \\ \hline
92 & -86 & 3 & -83 & 2,8 & -80 & 4 & -85 & 2,7 \\ \hline
93 & -81 & 5,2 & -85 & 2,7 & -80 & 6,2 & -75 & 4,4 \\ \hline
94 & -82 & 3,5 & -87 & 3,6 & -83 & 4,2 & -83 & 3,2 \\ \hline
95 & -81 & 4,6 & -79 & 2,2 & -72 & 3,7 & -80 & 6,1 \\ \hline
96 & -75 & 2,5 & -79 & 2,2 & -69 & 1 & -73 & 5,4 \\ \hline
97 & -78 & 5,8 & -70 & 3,5 & -68 & 3,2 & -73 & 5,9 \\ \hline
98 & -82 & 3,7 & -75 & 4,7 & -73 & 4,5 & -80 & 1,7 \\ \hline
99 & -80 & 3,2 & -76 & 2,5 & -77 & 3,6 & -80 & 5,4 \\ \hline
100 & -83 & 4,2 & -78 & 4,6 & -84 & 5,4 & -78 & 2,2 \\ \hline
101 & -77 & 4 & -78 & 4,6 & -76 & 2,5 & -77 & 4,1 \\ \hline
102 & -74 & 3,1 & -78 & 4,5 & -70 & 3,6 & -72 & 1,4 \\ \hline
103 & -74 & 3,7 & -77 & 5,6 & -75 & 1,4 & -78 & 2,7 \\ \hline
104 & -66 & 1,4 & -58 & 1 & -75 & 5,7 & -77 & 3,9 \\ \hline
105 & -77 & 5,7 & -78 & 2 & -75 & 1,7 & -83 & 5,7 \\ \hline
106 & -79 & 3,6 & -77 & 3,1 & -82 & 4 & -87 & 2,7 \\ \hline
107 & -81 & 4,2 & -78 & 2 & -84 & 3,7 & -85 & 3,9 \\ \hline
108 & -82 & 4,5 & -83 & 5 & -81 & 4 & -81 & 4,1 \\ \hline
109 & -84 & 2,5 & -82 & 4,1 & -83 & 4,1 & -84 & 4,7 \\ \hline
110 & -83 & 4,4 & -82 & 4,8 & -84 & 3,2 & -81 & 3,6 \\ \hline
111 & -83 & 4,3 & -80 & 4,8 & -79 & 3,2 & -83 & 3,6 \\ \hline
112 & -83 & 4 & -78 & 1 & -78 & 3,7 & -84 & 2,8 \\ \hline
113 & -79 & 3 & -75 & 2,5 & -82 & 5,4 & -86 & 3 \\ \hline
114 & -81 & 3,5 & -82 & 2,2 & -79 & 4 & -83 & 3,5 \\ \hline
115 & -87 & 2,8 & -82 & 4,3 & -82 & 2,5 & -88 & 3,5 \\ \hline
116 & -81 & 5,5 & -80 & 4 & -81 & 6,1 & -84 & 2,2 \\ \hline
117 & -83 & 2,7 & -83 & 3,5 & -85 & 3,3 & -86 & 3,9 \\ \hline
118 & -85 & 3,3 & -84 & 2 & -86 & 2,7 & -84 & 3 \\ \hline
119 & -87 & 2,5 & -80 & 2,8 & -79 & 3,2 & -89 & 1 \\ \hline
120 & -87 & 2,8 & -76 & 1,4 & -77 & 3 & -83 & 3,5 \\ \hline
121 & -83 & 4,1 & -81 & 4,6 & -75 & 1,7 & -82 & 2,2 \\ \hline
122 & -85 & 3,9 & -79 & 4,5 & -80 & 3,6 & -84 & 5,2 \\ \hline
123 & -79 & 1,7 & -76 & 3,2 & -77 & 2,7 & -81 & 3 \\ \hline
\end{tabular}

\caption{RSS values fluctuated heavily based on the orientation of the user. Sampling points referred to in the table are depicted in figure 6.}
\end{table}









\begin{figure}[h!]
\begin{center}
\includegraphics[width=\textwidth]{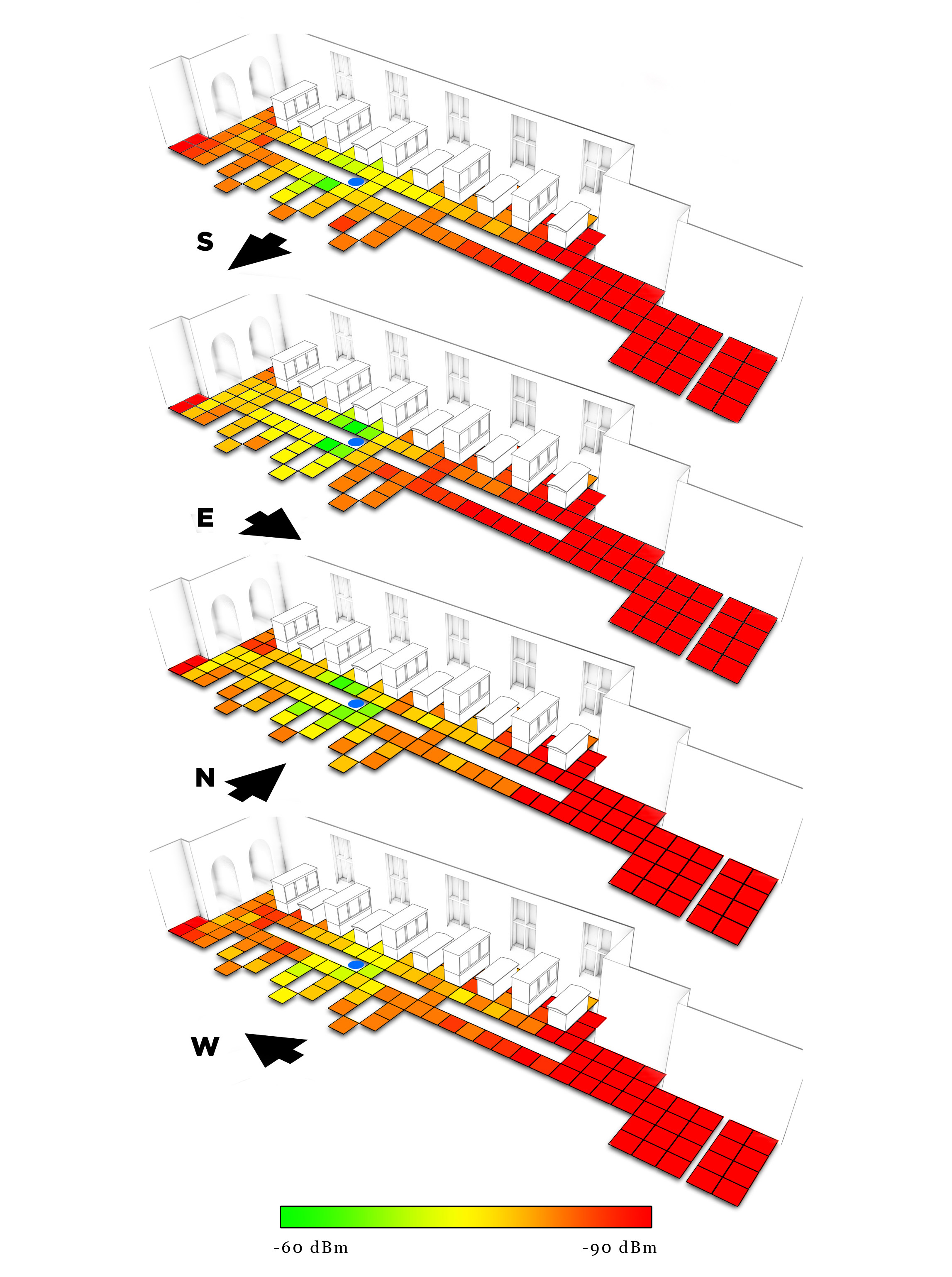}
\caption{The direction the user was facing had a dramatic impact on RSS.}
\end{center}
\end{figure}

\begin{figure}[h!]
\begin{center}
\includegraphics[width=\textwidth]{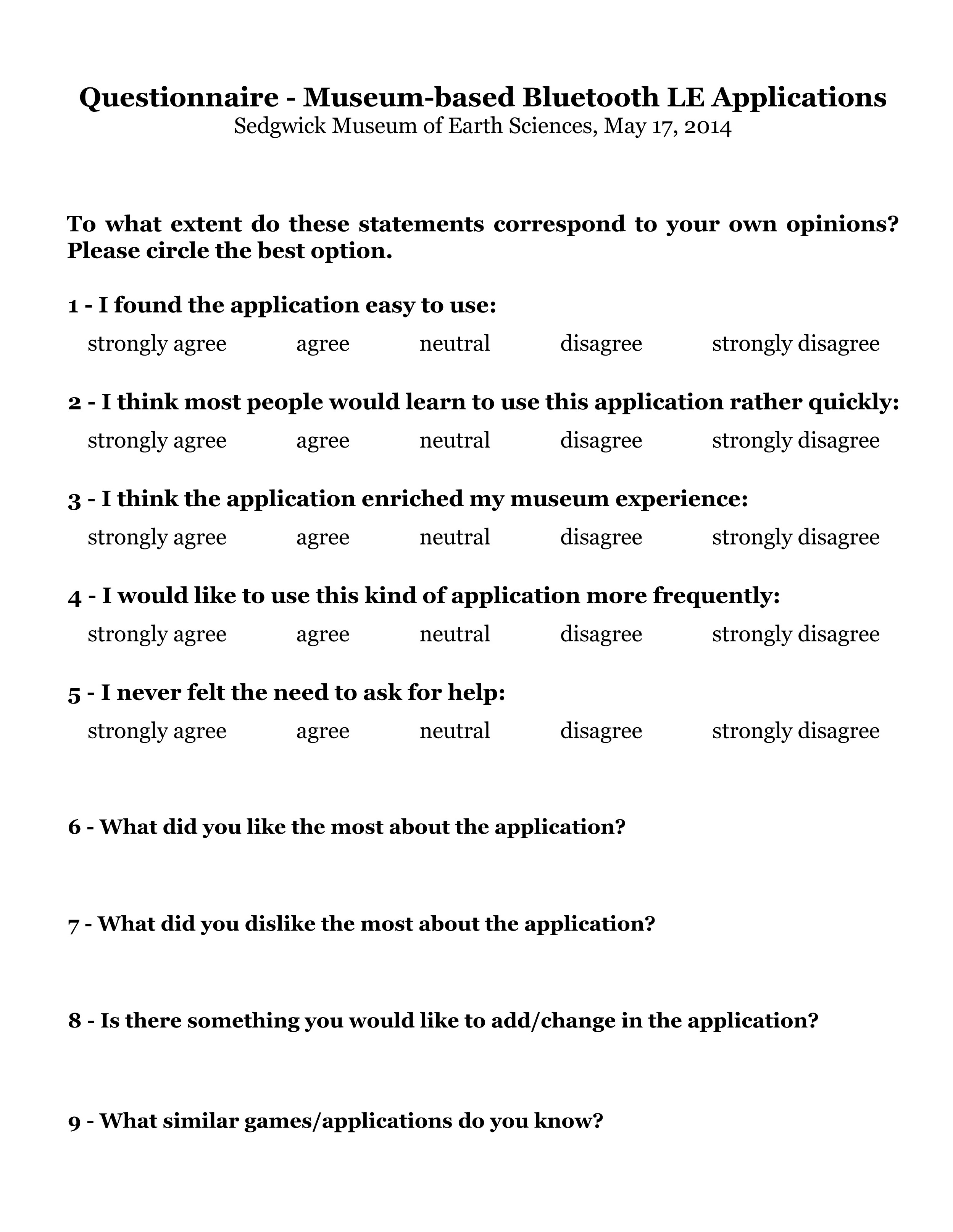}
\caption{A snapshot of the questionnaire used during our formative evaluation.}
\end{center}
\end{figure}



\received{Month Year}{Month Year}{Month Year}








\end{document}